\documentclass[usegraphicx,usenatbib]{mn2e}
\usepackage{times}
\def\apj {ApJ}
\def\apjl {ApJL}
\def\apjs {ApJS}
\def\aj {AJ}
\def\aap {A\&A}
\def\mnras {MNRAS}
\begin{document}
\title[Bars \& environment]
{Relating bars with the environment in the nearby Universe}
\author[H.J. Mart\'{\i}nez \& H. Muriel]
{H\'ector J. Mart\'{\i}nez\thanks{E-mail: julian@oac.uncor.edu}
\& Hern\'an Muriel\\
Instituto de Astronom\'{\i}a Te\'orica y Experimental,
IATE, CONICET$-$Observatorio Astron\'omico, Universidad Nacional de C\'ordoba,\\
Laprida 854, X5000BGR, C\'ordoba, Argentina \\
}
\date{\today}
\pagerange{\pageref{firstpage}--\pageref{lastpage}} 
\maketitle
\label{firstpage}
%%%%%%%%%%%%%%%%%%%%%%%%%%%%%%%%%%%%%%%%%%%%%%%%%%%%%%%%%%%%%%%%%%%%%%%%%%%%%%%
%%%%%%%%%%%%%%%%%%%%%%%%%%%%%%%%%%%%%%%%%%%%%%%%%%%%%%%%%%%%%%%%%%%%%%%%%%%%%%%
\begin{abstract}
We study the correlation between the fraction of barred spiral galaxies 
and environmental parameters of galaxies to understand in which environments
the bars are more commonly found. For this purpose we apply the Blanton et al. technique
to a sample of spiral galaxies drawn from the Nair \& Abraham catalogue. 
Our results agree with previous findings in which the fraction of barred galaxies is
almost insensitive to environment.
\end{abstract}
%%%%%%%%%%%%%%%%%%%%%%%%%%%%%%%%%%%%%%%%%%%%%%%%%%%%%%%%%%%%%%%%%%%%%%%%%%%%%%%
%%%%%%%%%%%%%%%%%%%%%%%%%%%%%%%%%%%%%%%%%%%%%%%%%%%%%%%%%%%%%%%%%%%%%%%%%%%%%%%
\section{Introduction} 
Bars are believed to be closely related with the dynamical evolution of disk galaxies and 
to play an important role in redistributing the angular momentum between dark and baryonic 
matter \citep{weinberg85,debattista98}. \citet{athanassoula03} proposed that this exchange of angular 
momentum is closely related to the density and velocity dispersion of the host halo. 
Other roles that have been assigned to bars are: 
to transport material to the centre and ignite starbursts 
\citep{sheth05} and/or feed the central black hole \citep{shlosman90,corsini03},
however no direct evidence of this is seen \citep{mulchaey97}; change chemical 
abundance gradient \citep{zaritsky92,martin94}; bars can trigger 
star formation along themselves, or have a lot of gas and no star formation 
\citep{kenney94,sheth02}. Bars are also associated with circumnuclear 
star formation activity \citep{sersic67,ho97,sheth00}; the formation of nuclear pseudo-bulges 
\citep{kormendy82,kormendy04}, rings \citep{schwarz81,buta86,buta96} and, possibly, spiral arms 
\citep{lindblad60,elmegreen85}.

For a fixed circular velocity, barred and unbarred galaxies have similar properties like 
luminosity, scale lengths, star formation  and colour \citep{courteau03}, which does not necessary imply
they have followed the same evolutionary paths (see \citealt{sheth08}). 
\citet{devaucouleurs63} using blue plates found that more than 60\% of nearby galaxies are barred. 
Similar fractions are observed in the near-infrared \citep{eskridge00,laurikainen04,menendez07}. 
In numerical simulations, bars appear naturally once the cold and rotationally supported disk is in 
place (see \citealt{athanassoula05} for a review), 
although, the precise mechanisms that drive this phenomenon are 
not well established yet. Among the internal mechanisms that can produce bars is the instability of 
the disk \citep{sellwood93,heller07,athanassoula08}.  

It has also been suggested that environment could also be important in leading to the formation of 
bars, although results are contradictory. 
In some numerical simulations bars are created in mergers and interactions 
\citep{walker96,mihos97,berentzen04}, it is transient in others \citep{gerin90} 
and in some others bars are destroyed as the galaxy becomes an elliptical. 
\citet{elmegreen90} analyse galaxies in binary and group systems and in the field 
and find a correlation between the bar fraction and environment for early-type spirals, with the highest 
fraction corresponding to binary systems. On the other hand, \citet{vandenbergh02} use 930 galaxies 
from the Palomar Sky Survey and conclude that the bar fraction does not depend on the environment. 
An environment that has been particularly studied is that of galaxy clusters. 
\citet{mendez10} study both, 
the centre and the infall regions of the Coma Cluster, in a wide range of magnitudes. They 
find that bars are hosted in galaxies in a tight interval of masses and luminosities 
($10^{9}\le M_*/M_{\odot}\le 10^{11}$ and $-22<M_r<-17$ respectively). These authors do
not find a significant difference in the fraction of bars between galaxies in the centre and 
in the infall regions suggesting that the cluster environment plays a second-order role in the 
bar formation/evolution.

The purpose of this paper is to evaluate different parameters that 
characterise the environment as possible generators of bars. We use the 
technique proposed by \citet{blanton05} and significance criteria introduced 
by \citet{martinez08} to samples of galaxies drawn from the catalogue by \citet{nair10a} (hereafter NA10). 
The paper is organised as follows: section 2 describe the sample of galaxies
and the environmental properties we use throughout our work; section 3 presents 
the results; finally we provide a discussion of our findings in section 4.
Throughout this paper we assume a flat cosmological model with parameters $H_0=70{\rm km~s^{-1}Mpc^{-1}}$,
$\Omega_M=0.3$ and $\Omega_{\Lambda}=0.7$.
%%%%%%%%%%%%%%%%%%%%%%%%%%%%%%%%%%%%%%%%%%%%%%%%%%%%%%%%%%%%%%%%%%%%%%%%%%%%%%%
%%%%%%%%%%%%%%%%%%%%%%%%%%%%%%%%%%%%%%%%%%%%%%%%%%%%%%%%%%%%%%%%%%%%%%%%%%%%%%%
\section{The galaxy sample}
For the purposes of this paper we use a sample of galaxies drawn from the
catalogue by NA10. This catalogue presents detailed visual classification
for 14034 galaxies in the Main Galaxy Sample \citep[MGS;][]{mgs} of the Fourth Data Release of the 
SDSS \citep[DR4;][]{dr4} that constitute a complete sample with $0.01\le z\le 0.1$ and down to
a limiting extinction corrected apparent magnitude of $g=16$. Each galaxy in the catalogue has
been morphologically classified by NA10 into T-Types. Additionally, they recorded the existence of
structures such as bars, rings, lenses, tails, warps, etc.

In our analyses below, we include all galaxies classified as spirals (T-Type$\ge 0$) in the 
NA10 catalogue which have axial ratio $b/a>0.55$. Below this cut-off the fraction of barred
galaxies drops dramatically (fig. 21 in NA10). There are no other important incompleteness
regarding bar identifications in the NA10. Our sample has 5508 galaxies, among which
1841 are barred. 

For the analysis of the correlation between the existence of bars and the environment
in which galaxies are located, we selected 4 measures of environment quoted in the NA10 catalogue
and also computed the projected distance to the nearest MGS neighbour
brighter than $M_r=-20$ (that is, a volume limited sample of galaxies up to $z=0.1$) and with 
$c|\Delta z| \le 1000{\rm km~s^{-1}}$ as 
another environment measure:
\begin{enumerate}
\item $L_{{\rm group}},g$: group luminosity ($9.5\le \log(L_{{\rm group},g}/L_{\odot g})\le 12.5$) from
\citet{yang07}.
\item $M_{\rm group}$: group mass ($9.5\le \log(M_{\rm group}/M_{\odot})\le 13.0$) from
\citet{yang07}.
\item $M_{\rm halo}$: group halo mass ($11.5\le \log(M_{\rm halo}/M_{\odot})\le 15.5$)
from \citet{yang07}.
\item $\Sigma_5$: 5th neighbour projected density ($-0.8\le \log(\Sigma_5/{\rm Mpc^{-2}})\le 1.8$)
from \citet{baldry06}.
\item $r_{NN}$: projected distance to the nearest neighbour ($0\le r_{NN}/{\rm kpc}\le 1500$) computed
in this work. We have not introduced corrections accounting for the well known fiber collision
incompleteness of the MGS, nevertheless, this should not bias our results since, both, barred
and non-barred spirals should be affected in the same way.
\end{enumerate}
All cut-offs in the lists above were imposed as a compromise between probing the largest 
possible volume in the space of parameters and, at the same time, having enough galaxies 
per bin to properly
carry out the correlation analyses detailed in the next section.

%%%%%%%%%%%%%%%%%%%%%%%%%%%%%%%%%%%%%%%%%%%%%%%%%%%%%%%%%%%%%%%%%%%%%%%%%%%%%%%
%%%%%%%%%%%%%%%%%%%%%%%%%%%%%%%%%%%%%%%%%%%%%%%%%%%%%%%%%%%%%%%%%%%%%%%%%%%%%%%
\section{Relating bars with environments}
We explore the ability of different environment parameters to predict the
fraction of barred galaxies by using the $\sigma_X$ statistics as defined by \citet{blanton05}. 
Briefly, given a set of environmental properties, $X_1,...,X_N$, the fraction of barred
galaxies, $f_{\rm bar}$, will be, in principle, a function of them: 
$f_{\rm bar}=f_{\rm bar}(X_1,...,X_N)$.
If we consider now a particular parameter $X_I$ and marginalise $f_{\rm bar}$ over the remaining ones,
we get the fraction of barred  galaxies as a function of $X_I$ alone, $f_{\rm bar}(X_I)$. 
The parameter $X_J$ that correlates best with the fraction of barred galaxies
will be the one that minimises the variance $\sigma_{X_J}$ of $f_{\rm bar}(X_J)$ after 
subtracting its global trend (see for details \citealt{blanton05,hm2,martinez08}). 

%%%%%%%%%%%%%%%%%%%%%%%%%%%%%%%%%%%%%%%%%%%%%%%%%%%%%%%%%%%%%%%%%%%%%%%%%%%%%%%
%%%%%%%%%%%%%%%%%%%%%%%%%%%%%%%%%%%%%%%%%%%%%%%%%%%%%%%%%%%%%%%%%%%%%%%%%%%%%%%
\begin{figure}
\includegraphics[width=90mm]{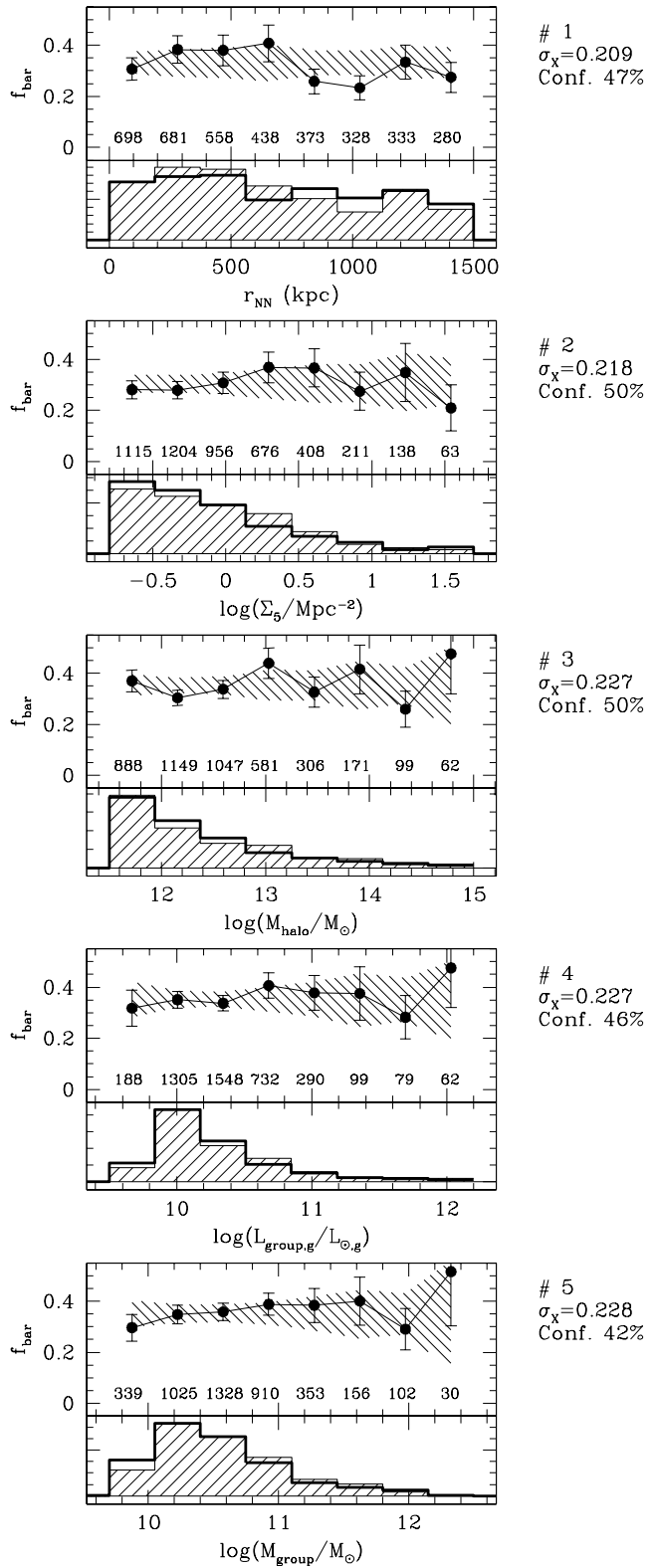}
\caption{The fraction of barred galaxies as a function of different measures of the 
environment, quoted error-bars are obtained by the bootstrap resampling technique.
From {\em upper} to {\em lower}, panels are sorted according to increasing $\sigma_X$ values. 
{\em Shaded areas} are the mean values and $\pm1\sigma$ error-bars from the resamplings.
We quote in the {\em bottom} the number of galaxies in each bin.
Below each panel we show the distributions corresponding to barred ({\em shaded
histogram}) and non-barred galaxies ({\em empty histogram}), both normalised to have the same area.}
\label{fig1}
\end{figure}
%%%%%%%%%%%%%%%%%%%%%%%%%%%%%%%%%%%%%%%%%%%%%%%%%%%%%%%%%%%%%%%%%%%%%%%%%%%%%%%
%%%%%%%%%%%%%%%%%%%%%%%%%%%%%%%%%%%%%%%%%%%%%%%%%%%%%%%%%%%%%%%%%%%%%%%%%%%%%%%

The $\sigma_X$ statistics provides the environmental parameter that predicts best the presence
of bars. This does not mean, however, that the best ranked parameter is a good predictor
of bars. To complement the $\sigma_X$ statistics, we use the significance criterion
by \citet{martinez08} computed by using bootstrap resamplings.

Since we are dealing with a flux-limited galaxy sample in our analyses, we have 
weighted each galaxy in our statistics by $1/V_{\rm max}$ \citep{schmidt68}. 
Given that the fraction of barred galaxies depends on absolute magnitude
\citep{sheth08,nair10b}, and that brighter galaxies tend to be located in higher density
environments, the non inclusion of such a weighting scheme can lead to systematics 
in our analyses. 

In Fig. \ref{fig1} we show the results of our analysis on the relationship between 
bars and environment. Panels are sorted from top to bottom according to 
increasing $\sigma_X$ values.
We quote next to each panel the $\sigma_X$ value along with its significance. 
Each panel shows the fraction, $f_{\rm bar}(X)$, of barred galaxies as a function of the corresponding
quantity $X$. Error bars are 1 $\sigma$ bootstrap resampling error-bars. Shaded areas
represent the overall mean value of $f_{\rm bar}(X)$ plus/minus 1 $\sigma$ from the bootstrap resamplings.
We also show in the inferior part of each panel the $1/V_{\rm max}$ weighted histograms
of barred (shaded) and non-barred (empty) galaxies, all these histograms are normalised
to have the same area.  
The best predicting parameter is the distance to the nearest neighbour, in the sense that
galaxies with bars tend to have closer neighbours. This is followed by 
the projected galaxy density, the mass of the halo, the group luminosity and finally
the group mass. Nevertheless, all the parameters have significance below 68\%, i.e., 
less that 1$\sigma$ for Gaussian statistics, thus none of them is significant according to
our criterion.

Despite that none of the explored parameters are closely correlated to the presence
of bars, the trends in Fig. \ref{fig1} suggest that barred galaxies are slightly
more common at higher densities, and in more massive halos.
Inspired in the results by \citet{elmegreen90}, we also check whether the correlation
of bars and the environment is stronger for early spirals, by repeating the $\sigma_X$
computations only for galaxies classified as Sa, Sab and Sb by NA10. We find no evidence
of significant correlation for those galaxies either. 
The results suggest, at best, a second order environmental effect.

We have repeated our computations without the $1/V_{\rm max}$ weighting scheme, 
resulting in a different ranking of parameters. In this case we observe that:
(i) the overall $f_{\rm bar}$ is slightly higher since $f_{\rm bar}$ is an increasing 
function of luminosity;
(ii) the trends as in Fig. \ref{fig1} do not qualitatively change, however, (iii)
the variances around the global trends, $\sigma_X$, 
do change leading to a different ranking: the group mass ranks first and the remaining
four parameters have almost identical $\sigma_X$ values. This is no surprise since 
in this case the statistics is dominated by brighter galaxies, that tend to be located in 
more massive systems, and among them the fraction of bars is higher. 
Again, we find that significance levels are well below 68\%.
The non inclusion of a weighting scheme accounting for the fact that
we are dealing with an essentially flux limited galaxy sample can lead to wrong conclusions.

Although non-significant according to our criterion, the distance to the nearest 
neighbour is the environmental parameter that correlates best with the fraction of bars. 
The nearest neighbour in our analysis is, in all cases, a relatively bright object,
$M_r<-20$, since we searched for it in a volume limited (up to $z=0.1$) sample of galaxies drawn from the 
MGS to avoid biases with redshift. If the proximity to another galaxy is an important
factor for the formation and/or stability of a bar, the mass ratio between a galaxy and its closest
neighbour might be important as well.
We have further explored the correlation between bars and neighbours 
in terms of the difference in absolute magnitude between the barred galaxies 
and their nearest neighbours, which is broadly related to their mass ratio.
We classified spiral galaxies in the NA10 catalogue according to the $r-$band absolute magnitude
difference with the nearest neighbour, $\Delta M=M_{NN}-M$, into two groups: galaxies whose nearest 
neighbour is brighter or have similar luminosity ($\Delta M\le 0.5$); and galaxies whose nearest 
neighbour is fainter ($1.5>\Delta M > 0.5$). 
In this case, we searched for the nearest neighbours in the flux limited ($r\le 17.77$)
MGS, and restricted the sample of spiral galaxies in the NA10 to those that are at least
1.5 magnitudes brighter than the absolute magnitude corresponding to the apparent magnitude
limit at the galaxy's redshift.
For these two subsamples we have studied the trend of the fraction of barred galaxies as
a function of the projected distance to the nearest neighbour, as in the {\em upper panel}
of Fig. \ref{fig1}, and have found no significant differences.
Thus, according to our results, bars are located in galaxies with no particular
preference regarding the luminosity of the nearest neighbour.

%%%%%%%%%%%%%%%%%%%%%%%%%%%%%%%%%%%%%%%%%%%%%%%%%%%%%%%%%%%%%%%%%%%%%%%%%%%%%%%
%%%%%%%%%%%%%%%%%%%%%%%%%%%%%%%%%%%%%%%%%%%%%%%%%%%%%%%%%%%%%%%%%%%%%%%%%%%%%%%

\section{Conclusions}
We have studied the relationship between the fraction of barred spiral galaxies
and a number of environmental parameters in the nearby universe using a
complete sample of spiral galaxies taken from the NA10 catalogue. 
For this purpose we have applied the technique by \citet{blanton05} and the significance 
criterion by \citet{martinez08}, to a sample of spiral galaxies taken from the NA10
catalogue. Once the range of a set of parameters is defined, this technique measures the 
ability of each parameter to predict an observable, in this case, the bar fraction.

We have considered in our analysis 5 parameters characterising the environment:
group luminosity, group mass, group halo mass (the three of them computed by
\citealt{yang07}), the projected galaxy density (computed by \citealt{baldry06})
and the projected distance to the nearest neighbour (computed in this work).
Our results indicate that the latter parameter is the one that best predicts the
existence of bars, however the signal is only marginal. 
The small effect is in the sense that spiral galaxies with a
nearest neighbour within 0.5 Mpc tend to have a slightly higher fraction of bars.
We find no evidence of bars preferring systematically brighter or fainter neighbours.
Our finding that the proximity to another galaxy could play a role
in the formation of a bar, can be related to the predictions of some numerical simulations in which 
interactions can trigger bars and/or influence bar properties 
(e.g. \citealt{noguchi87,gerin90,mihos97,miwa98,berentzen04}).
However, due to the low significance we find, our results can also be in agreement with
\citet{vandenbergh02} and \citet{mendez10}, in which the fraction
of barred galaxies does not depend on the environment.
Numerical simulations have shown that bars form naturally in discs 
(e.g. \citealt{athanassoula05}
and references therein), 
thus, bars could be understood in terms of nature only, leaving for nurture, at best, 
a secondary role.  A larger sample of barred galaxies is needed to shed more light on the 
relationship between environment and the presence of bars.

According to our results, the relation between environment and bars appears 
to be, at best, a second order effect. Since the NA10 sample contains galaxies
in a wide range of environments (see histograms in Fig. \ref{fig1}), it is worth emphasising 
the fact that the presence of bars do not seem to depend on environment while
most galaxy properties do. 

%%%%%%%%%%%%%%%%%%%%%%%%%%%%%%%%%%%%%%%%%%%%%%%%%%%%%%%%%%%%%%%%%%%%%%%%%%%%%%%
%%%%%%%%%%%%%%%%%%%%%%%%%%%%%%%%%%%%%%%%%%%%%%%%%%%%%%%%%%%%%%%%%%%%%%%%%%%%%%%
\section*{Acknowledgements}
This work has been partially supported with grants from Consejo Nacional
de Investigaciones Cient\'\i ficas y T\'ecnicas de la Rep\'ublica Argentina
(CONICET), Secretar\'\i a de Ciencia y Tecnolog\'\i a de la Universidad 
de C\'ordoba and Ministerio de Ciencia y Tecnolog\'ia de la Provincia de C\'ordoba, Argentina.

Funding for the SDSS and SDSS-II has been provided by the Alfred P. Sloan Foundation, the Participating Institutions, the NSF, the U.S. Department of Energy, NASA, the Japanese Monbukagakusho, the Max Planck Society, and the Higher Education Funding Council for England. The SDSS Web Site is http://www.sdss.org/.
The SDSS is managed by the Astrophysical Research Consortium for the Participating Institutions. The Participating Institutions are the American Museum of Natural History, Astrophysical Inst. Potsdam, Univ. of Basel, Univ. of Cambridge, Case Western Reserve Univ., Univ. of Chicago, Drexel Univ., Fermilab, the Inst. for Advanced Study, the Japan Participation Group, Johns Hopkins Univ., the Joint Inst. for Nuclear Astrophysics, the Kavli Inst. for Particle Astrophysics and Cosmology, the Korean Scientist Group, the Chinese Academy of Sciences (LAMOST), Los Alamos National Laboratory, the Max-Planck-Inst. for Astronomy (MPIA), the Max-Planck-Inst. for Astrophysics (MPA), New Mexico State Univ., Ohio State Univ., Univ. of Pittsburgh, Univ. of Portsmouth, Princeton Univ., the United States Naval Observatory, and the Univ. of Washington.
%%%%%%%%%%%%%%%%%%%%%%%%%%%%%%%%%%%%%%%%%%%%%%%%%%%%%%%%%%%%%%%%%%%%%%%%%%%%%%%
%%%%%%%%%%%%%%%%%%%%%%%%%%%%%%%%%%%%%%%%%%%%%%%%%%%%%%%%%%%%%%%%%%%%%%%%%%%%%%%

\label{lastpage}
\end{document}